\documentclass[journal]{IEEEtran}
\usepackage[ruled,vlined,linesnumbered]{algorithm2e}
\usepackage{verbatim}
\usepackage{amsfonts}
\usepackage{amssymb}
\usepackage{stfloats}
\usepackage{cite}
\usepackage{graphicx}
\usepackage{psfrag}
\usepackage{amsmath}
\usepackage{array}
\usepackage{epstopdf}
\usepackage{authblk}
\usepackage{graphicx} 
\usepackage{amsthm} 
\usepackage{lipsum}
\usepackage{verbatim} 
\usepackage{authblk}
\usepackage{mathtools}
\usepackage{cuted}
\usepackage{booktabs}
\usepackage{bm}
\allowdisplaybreaks[2]
\usepackage{amsmath}

\usepackage{array}

\usepackage{url}
\usepackage{graphicx,amsmath,amssymb,amsfonts}

\usepackage{graphicx}  
\usepackage{float}  
\usepackage{subfigure}  
\usepackage[colorlinks=true]{hyperref}
\hypersetup{
	citecolor = {blue},
}

\usepackage{xcolor}

\makeatletter

\newcommand{\Rmnum}[1]{\expandafter\@slowromancap\romannumeral #1@}
\makeatother

\usepackage{cancel}  

\usepackage{lipsum}

\begin{document}
	% reference control
	\bstctlcite{ref:BSTcontrol}
	\newtheorem{theorem}{\bf Theorem}
\newtheorem{acknowledgement}[theorem]{Acknowledgement}
\newtheorem{axiom}[theorem]{Axiom}
\newtheorem{case}[theorem]{Case}
\newtheorem{claim}[theorem]{Claim}
\newtheorem{conclusion}[theorem]{Conclusion}
\newtheorem{condition}[theorem]{Condition}
\newtheorem{conjecture}[theorem]{Conjecture}
\newtheorem{criterion}[theorem]{Criterion}
\newtheorem{definition}{\bf Definition}
\newtheorem{exercise}[theorem]{Exercise}
\newtheorem{lemma}{\bf Lemma}
\newtheorem{corollary}{\bf Corollary}
\newtheorem{notation}[theorem]{Notation}
\newtheorem{problem}[theorem]{Problem}
\newtheorem{proposition}{\bf Proposition}
\newtheorem{solution}[theorem]{Solution}
\newtheorem{summary}[theorem]{Summary}
\newtheorem{assumption}{\bf Assumption}
\newtheorem{example}{\bf Example}
\newtheorem{remark}{\bf Remark}
\newtheorem{granularity_constraint}[theorem]{Granularity Constraint}

\def\qed{$\blacksquare$}
\def\QED{\mbox{\phantom{m}}\nolinebreak\hfill$\,\blacksquare$}
\def\proof{\noindent{\emph{Proof:} }}
\def\poof{\noindent{\emph{Sketch of Proof:} }}
\def
\endproof{\hspace*{\fill}~\qed
\par
\endtrivlist\unskip}
\def\endproof{\hspace*{\fill}~\qed\par\endtrivlist\vskip3pt}

\def\E{\mathsf{E}}
\def\eps{\varepsilon}
\def\phi{\varphi}
\def\Lsp{{\boldsymbol L}}
\def\Bsp{{\boldsymbol B}}
\def\lsp{{\boldsymbol\ell}}
\def\Ltsp{{\Lsp^2}}
\def\Lpsp{{\Lsp^p}}
\def\Linsp{{\Lsp^{\infty}}}
\def\LtR{{\Lsp^2(\Rst)}}
\def\ltZ{{\lsp^2(\Zst)}}
\def\ltsp{{\lsp^2}}
\def\ltZt{{\lsp^2(\Zst^{2})}}
\def\ninN{{n{\in}\Nst}}
\def\oh{{\frac{1}{2}}}
\def\grass{{\cal G}}
\def\ord{{\cal O}}
\def\dist{{d_G}}
\def\conj#1{{\overline#1}}
\def\ntoinf{{n \rightarrow \infty }}
\def\toinf{{\rightarrow \infty }}
\def\tozero{{\rightarrow 0 }}
\def\trace{{\operatorname{trace}}}
\def\ord{{\cal O}}
\def\UU{{\cal U}}
\def\rank{{\operatorname{rank}}}
\def\acos{{\operatorname{acos}}}

\def\SINR{\mathsf{SINR}}
\def\SNR{\mathsf{SNR}}
\def\SIR{\mathsf{SIR}}
\def\tSIR{\widetilde{\mathsf{SIR}}}
\def\Ei{\mathsf{Ei}}
\def\l{\left}
\def\r{\right}
\def\({\left(}
\def\){\right)}
\def\lb{\left\{}
\def\rb{\right\}}

\setcounter{page}{1}

% Definitions
\newcommand{\eref}[1]{(\ref{#1})}
\newcommand{\fig}[1]{Fig.\ \ref{#1}}
\newcommand{\Stage}[1]{\tcp*{\textbf{#1}}}

% Bold lowercase
\def\bydef{:=}
\def\ba{{\mathbf{a}}}
\def\bb{{\mathbf{b}}}
\def\bc{{\mathbf{c}}}
\def\bd{{\mathbf{d}}}
\def\bee{{\mathbf{e}}}
\def\bff{{\mathbf{f}}}
\def\bg{{\mathbf{g}}}
\def\bh{{\mathbf{h}}}
\def\bi{{\mathbf{i}}}
\def\bj{{\mathbf{j}}}
\def\bk{{\mathbf{k}}}
\def\bl{{\mathbf{l}}}
\def\bn{{\mathbf{n}}}
\def\bo{{\mathbf{o}}}
\def\bp{{\mathbf{p}}}
\def\bq{{\mathbf{q}}}
\def\br{{\mathbf{r}}}
\def\bs{{\mathbf{s}}}
\def\bt{{\mathbf{t}}}
\def\bu{{\mathbf{u}}}
\def\bv{{\mathbf{v}}}
\def\bw{{\mathbf{w}}}
\def\bx{{\mathbf{x}}}
\def\by{{\mathbf{y}}}
\def\bz{{\mathbf{z}}}
\def\b0{{\mathbf{0}}}

% Bold capital letters
\def\bA{{\mathbf{A}}}
\def\bB{{\mathbf{B}}}
\def\bC{{\mathbf{C}}}
\def\bD{{\mathbf{D}}}
\def\bE{{\mathbf{E}}}
\def\bF{{\mathbf{F}}}
\def\bG{{\mathbf{G}}}
\def\bH{{\mathbf{H}}}
\def\bI{{\mathbf{I}}}
\def\bJ{{\mathbf{J}}}
\def\bK{{\mathbf{K}}}
\def\bL{{\mathbf{L}}}
\def\bM{{\mathbf{M}}}
\def\bN{{\mathbf{N}}}
\def\bO{{\mathbf{O}}}
\def\bP{{\mathbf{P}}}
\def\bQ{{\mathbf{Q}}}
\def\bR{{\mathbf{R}}}
\def\bS{{\mathbf{S}}}
\def\bT{{\mathbf{T}}}
\def\bU{{\mathbf{U}}}
\def\bV{{\mathbf{V}}}
\def\bW{{\mathbf{W}}}
\def\bX{{\mathbf{X}}}
\def\bY{{\mathbf{Y}}}
\def\bZ{{\mathbf{Z}}}

% mathbb Bold capital letters
\def\mA{{\mathbb{A}}}
\def\mB{{\mathbb{B}}}
\def\mC{{\mathbb{C}}}
\def\mD{{\mathbb{D}}}
\def\mE{{\mathbb{E}}}
\def\mF{{\mathbb{F}}}
\def\mG{{\mathbb{G}}}
\def\mH{{\mathbb{H}}}
\def\mI{{\mathbb{I}}}
\def\mJ{{\mathbb{J}}}
\def\mK{{\mathbb{K}}}
\def\mL{{\mathbb{L}}}
\def\mM{{\mathbb{M}}}
\def\mN{{\mathbb{N}}}
\def\mO{{\mathbb{O}}}
\def\mP{{\mathbb{P}}}
\def\mQ{{\mathbb{Q}}}
\def\mR{{\mathbb{R}}}
\def\mS{{\mathbb{S}}}
\def\mT{{\mathbb{T}}}
\def\mU{{\mathbb{U}}}
\def\mV{{\mathbb{V}}}
\def\mW{{\mathbb{W}}}
\def\mX{{\mathbb{X}}}
\def\mY{{\mathbb{Y}}}
\def\mZ{{\mathbb{Z}}}

% Caligraphic capital letters
\def\cA{\mathcal{A}}
\def\cB{\mathcal{B}}
\def\cC{\mathcal{C}}
\def\cD{\mathcal{D}}
\def\cE{\mathcal{E}}
\def\cF{\mathcal{F}}
\def\cG{\mathcal{G}}
\def\cH{\mathcal{H}}
\def\cI{\mathcal{I}}
\def\cJ{\mathcal{J}}
\def\cK{\mathcal{K}}
\def\cL{\mathcal{L}}
\def\cM{\mathcal{M}}
\def\cN{\mathcal{N}}
\def\cO{\mathcal{O}}
\def\cP{\mathcal{P}}
\def\cQ{\mathcal{Q}}
\def\cR{\mathcal{R}}
\def\cS{\mathcal{S}}
\def\cT{\mathcal{T}}
\def\cU{\mathcal{U}}
\def\cV{\mathcal{V}}
\def\cW{\mathcal{W}}
\def\cX{\mathcal{X}}
\def\cY{\mathcal{Y}}
\def\cZ{\mathcal{Z}}
\def\cd{\mathcal{d}}
\def\Mt{M_{t}}
\def\Mr{M_{r}}
%% my defs
\def\O{\Omega_{M_{t}}}
\newcommand{\figref}[1]{{Fig.}~\ref{#1}}
\newcommand{\tabref}[1]{{Table}~\ref{#1}}

%% From Kaibin
\newcommand{\var}{\mathsf{var}}
\newcommand{\fb}{\tx{fb}}
\newcommand{\nf}{\tx{nf}}
\newcommand{\BC}{\tx{(bc)}}
\newcommand{\MAC}{\tx{(mac)}}
\newcommand{\Pout}{p_{\mathsf{out}}}
\newcommand{\nnn}{\nn\\}
\newcommand{\FB}{\tx{FB}}
\newcommand{\TX}{\tx{TX}}
\newcommand{\RX}{\tx{RX}}
\renewcommand{\mod}{\tx{mod}}
\newcommand{\m}[1]{\mathbf{#1}}
\newcommand{\td}[1]{\tilde{#1}}
\newcommand{\sbf}[1]{\scriptsize{\textbf{#1}}}
\newcommand{\stxt}[1]{\scriptsize{\textrm{#1}}}
\newcommand{\suml}[2]{\sum\limits_{#1}^{#2}}
\newcommand{\sumlk}{\sum\limits_{k=0}^{K-1}}
\newcommand{\eqhsp}{\hspace{10 pt}}
\newcommand{\tx}[1]{\texttt{#1}}
\newcommand{\Hz}{\ \tx{Hz}}
\newcommand{\sinc}{\tx{sinc}}
\newcommand{\tr}{\mathrm{tr}}
\newcommand{\diag}{\mathrm{diag}}
\newcommand{\MAI}{\tx{MAI}}
\newcommand{\ISI}{\tx{ISI}}
\newcommand{\IBI}{\tx{IBI}}
\newcommand{\CN}{\tx{CN}}
\newcommand{\CP}{\tx{CP}}
\newcommand{\ZP}{\tx{ZP}}
\newcommand{\ZF}{\tx{ZF}}
\newcommand{\SP}{\tx{SP}}
\newcommand{\MMSE}{\tx{MMSE}}
\newcommand{\MINF}{\tx{MINF}}
\newcommand{\RC}{\tx{MP}}
\newcommand{\MBER}{\tx{MBER}}
\newcommand{\MSNR}{\tx{MSNR}}
\newcommand{\MCAP}{\tx{MCAP}}
\newcommand{\vol}{\tx{vol}}
\newcommand{\ah}{\hat{g}}
\newcommand{\tg}{\tilde{g}}
\newcommand{\teta}{\tilde{\eta}}
\newcommand{\heta}{\hat{\eta}}
\newcommand{\uh}{\m{\hat{s}}}
\newcommand{\eh}{\m{\hat{\eta}}}
\newcommand{\hv}{\m{h}}
\newcommand{\hh}{\m{\hat{h}}}
\newcommand{\Po}{P_{\mathrm{out}}}
\newcommand{\Poh}{\hat{P}_{\mathrm{out}}}
\newcommand{\Ph}{\hat{\gamma}}
\newcommand{\mat}[1]{\begin{matrix}#1\end{matrix}}
\newcommand{\ud}{^{\dagger}}
\newcommand{\C}{\mathcal{C}}
\newcommand{\nn}{\nonumber}
\newcommand{\nInf}{U\rightarrow \infty}

%	\tiny, \scriptsize, \footnotesize, \small, \normalsize, \large, \Large, \LARGE, \huge, \Huge
	\title{\fontsize{22pt}{27pt}\selectfont Task-Oriented Multimodal Token Transmission in Resource-Constrained Multiuser Networks
	}

%\author{Authors}
\author{~Junhe~Zhang,~Wanli~Ni,~Pengwei~Wang,~Dongyu~Wang}
	
% make the title area
\maketitle

% two types footnote
\newcommand\blfootnote[1]{% 
	\begingroup 
	\renewcommand\thefootnote{}\footnote{#1}% 
	\addtocounter{footnote}{-1}% 
	\endgroup 
}

\blfootnote{
	This work of Junhe Zhang was supported in part by the China Scholarship Council (CSC).
	The work of Wanli Ni was supported in part by the Postdoctoral Fellowship Program of CPSF under Grant Number GZB20240386, and in part by the China Postdoctoral Science Foundation under Grant Number 2024M761669.
	\textit{(Corresponding author: Dongyu Wang.)}
	
	Junhe Zhang and Dongyu Wang are with the Key Laboratory
	of Universal Wireless Communication, Ministry of Education, Beijing University of Posts and Telecommunications, Beijing 100876, China (e-mail: zhangjunhe@bupt.edu.cn; dy\_wang@bupt.edu.cn).
	
	Wanli Ni is with the Department of Electronic Engineering, Tsinghua University, Beijing 100084, China (e-mail: niwanli@tsinghua.edu.cn).
	
	Pengwei Wang is with the Department of Mechanical Engineering, National University of Singapore, Singapore 117575, Singapore (email: wang\_pengwei@u.nus.edu).
	
	Code is available at: https://github.com/FlyingHat404/TokCom-MLMs.
}
\vspace{-4 mm}

\begin{abstract}
With the emergence of large model-based agents, widely adopted transformer-based architectures inevitably produce excessively long token embeddings for transmission, which may result in high bandwidth overhead, increased power consumption and latency.
In this letter, we propose a task-oriented multimodal token transmission scheme for efficient multimodal information fusion and utilization. To improve the efficiency of token transmission, we design a two-stage training algotithm, including cross-modal alignment and task-oriented fine-tuning, for large model-based token communication. Meanwhile, token compression is performed using a sliding window pooling operation to save communication resources.
To balance the trade-off between latency and model performance caused by compression, we formulate a weighted-sum optimization problem over latency and validation loss. We jointly optimizes bandwidth, power allocation, and token length across users by using an alternating optimization method.
Simulation results demonstrate that the proposed algorithm outperforms the baseline under different bandwidth and power budgets. Moreover, the two-stage training algorithm achieves higher accuracy across various signal-to-noise ratios than the method without cross-modal alignment.
\end{abstract}


\section{Introduction}
With the increasing demand for massive data transmission, semantic communication (SemCom), focusing on transmitting only task-relevant information, is envisioned as a core paradigm in 6G \cite{wanting-survey}.
To further harness vast amounts of multimodal data, the combination of semantic communication with multimodal large models (MLMs) has been explored recently, aiming to integrate diverse modalities into a unified framework \cite{multi-task}. Despite demonstrating strong feature processing capabilities, the fundamental component used in such systems, transformers, inevitably generates a large number of high-dimensional intermediate features, referred to as token embeddings.
Directly transmitting these token embeddings over wireless channels consumes substantial bandwidth and power resources, and significantly increases transmission latency, limiting its applicability in resource-constrained networks \cite{yuanwei}.

Fortunately, the inherent sequential structure and contextual dependencies of tokens make them highly amenable to compression. In \cite{token-prune}, the authors combined token pruning and merging to directly reduce token length, effectively lowering computational costs while preserving crucial information. In \cite{tokcom}, the authors proposed a token communication paradigm that exploits the contextual dependencies of token sequences for routing, prioritizing the transmission of less predictable tokens to reduce communication overhead. In \cite{embedding-com}, the authors proposed a multimodal token importance evaluation method that leverages inter-modal relationships and contextual dependencies to identify the most informative cross-modal tokens for transmission.
However, the above works either consider only a single modality or merely use cross-modal information as a basis for compression, without performing multimodal information fusion to enhance downstream task performance. 
Moreover, they do not explicitly model the impact of token compression on communication system performance, nor do they address the trade-off between resource savings and potential performance degradation caused by reducing tokens.

In this letter, we propose a task-oriented multimodal token transmission scheme that efficiently fuses multimodal information while conserving communication resources via token compression. We explicitly model the effects of token compression on transmission and model performance. Then, we formulate a optimization problem to balance latency and validation loss under constrained bandwidth and power. The contributions are summarized as follows.

\begin{itemize}
	\item
	To efficiently exploit multimodal information, we adopt a two-stage training algorithm consisting of cross-modal alignment and task-oriented fine-tuning. Specifically, a contrastive split fine-tuning method is first used to project heterogeneous modalities into a unified feature space, and then the entire system is fine-tuned on task-specific objectives to enhance task-relevant feature extraction.
	\item
	We formulate a weighted-sum optimization problem that jointly minimizes latency and validation loss. Then, we employ an alternating optimization (AO) approach to optimize bandwidth allocation, power control, and token compression across users.
	\item
	Simulation results reveal that the two-stage training algorithm outperforms the baseline (without cross-modal alignment), even when using fewer tokens. 
	For SNRs above 3 dB, the performance variation with the maximum token length remains within $0.5\%$, whereas halving the token length under low-SNR conditions leads to a $2.3\%$ performance degradation.
\end{itemize}

%\newpage
\section{Token Transmission Framework}

\begin{figure*}
	\centering
	\includegraphics[scale=0.38]{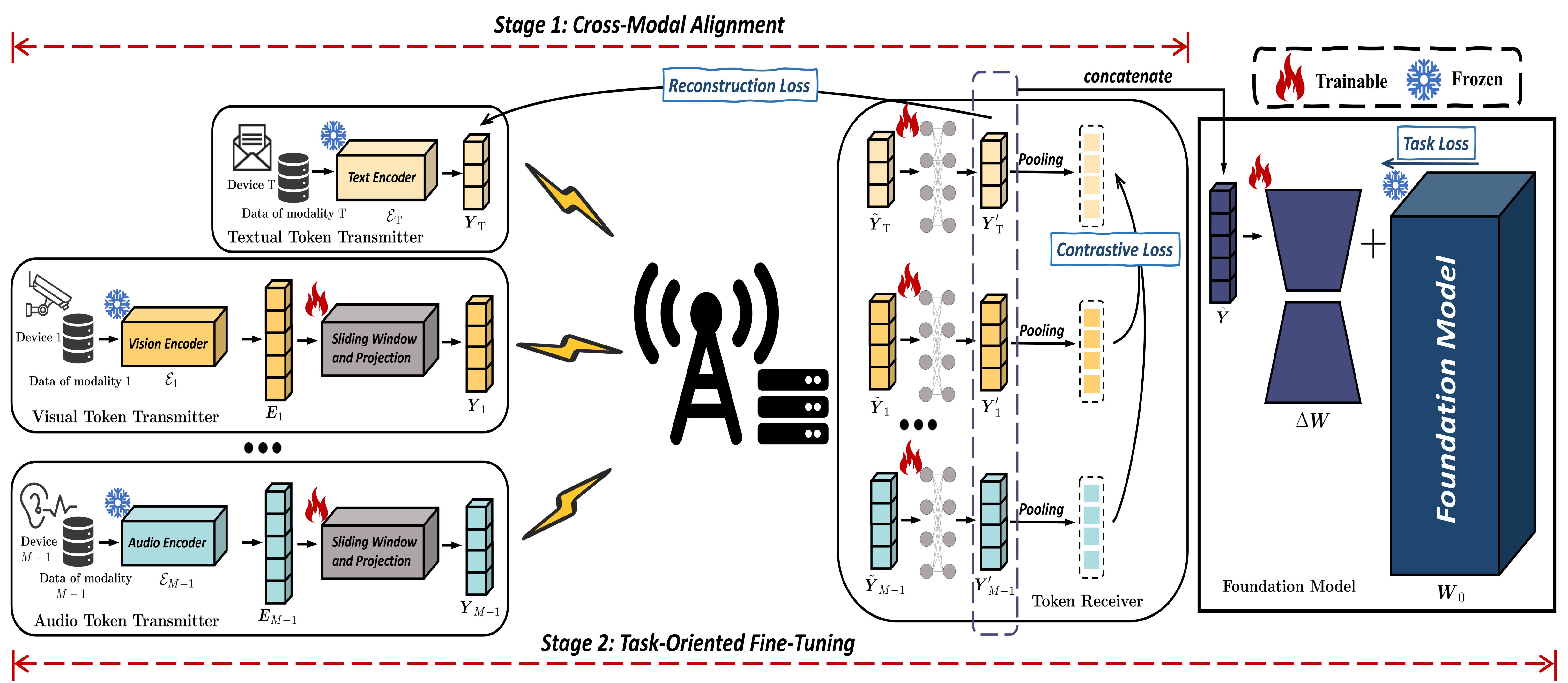}
	\caption{Task-oriented multimodal token transmission in resource-constrained multiuser networks.}
	\vspace{-3mm}
	\label{fig:SYS}
\end{figure*}

As illustrated in Fig.~\ref{fig:SYS}, we consider token transmission in a multiuser wireless network consisting of $M$ devices and a base station (BS) equipped with an edge server. The network handles $M$ modalities of data, with each device processing one modality.
For notational simplicity, we use the set $\mathcal{M}=\{1,2,\ldots,M-1, {\rm T}\}$ to represent the devices and corresponding modalities, with $\bar{\mathcal{M}}=\{1,2,\ldots,M-1\}$ denoting the non-text modalities and ${\rm T}$ denoting the text modality, so that $\mathcal{M}=\bar{\mathcal{M}}\cup{\rm T}$.
Token transmitters at the devices extract token embeddings from raw data, modulate and transmit them over the wireless channels. The received signals are processed at the BS by token receivers to recover token embeddings, which are then concatenated and forwarded to the foundation model deployed on the edge server for multimodal fusion and downstream task execution.

\subsection{Token Transmitter}
For text data $\bm{X}_{\rm T}$, the corresponding token $\bm{Y}_{\rm T}$ can be obtained as
\begin{align}\label{yt}
	\bm{Y}_{\rm T} = \mathcal{E}_{\rm T} (\bm{X}_{\rm T}),
\end{align}
where $\mathcal{E}_{\rm T}(\cdot)$ is the text encoder, and $\bm{Y}_{\rm T} \in \mathbb{R}^{s_{\rm T} \times d_{\rm T}}$ is with sequence length $s_{\rm T}$ and dimension $d_{\rm T}$.
For data $\bm{X}_{m}$ of modality $m \in \bar{\mathcal{M}}$, the corresponding token $\bm{Z}_{m}$ can be obtained as
\begin{align}\label{em}
	\bm{Z}_{m} = \mathcal{E}_{m} (\bm{X}_{m}), \quad \forall m \in \bar{\mathcal{M}},
\end{align}
where $\mathcal{E}_{m}(\cdot)$ is the encoder for modality $m$, and $\bm{Z}_{m} \in \mathbb{R}^{\hat{s}_{m} \times d_{m}}$ is with sequence length $\hat{s}_{m}$ and dimension $d_{m}$.
Non-textual modalities often produce token sequences with excessively long lengths $\hat{s}_{m}$. Directly transmitting $\bm{Z}_{m}$ over the wireless channel would therefore incur substantial bandwidth, power, and latency overhead. To reduce the token sequence length while retaining essential temporal patterns, we apply a sliding-window pooling approach. For each modality $m \in \bar{\mathcal{M}}$, the token sequence $\bm{Z}_{m}$ is divided into $s_m$ consecutive token sets, and a pooling operation is applied within each set to compress it into a single token. As a result, the original token sequence $\bm{Z}_{m}$ of length $\hat{s}_{m}$ is compressed into a shorter sequence of length $s_m$. The compressed token sequence is then passed through a fully connected layer to map its original feature dimension $d_m$ to the text token dimension $d_{\rm T}$, enabling subsequent integration into the foundation model. The entire process can be expressed as
\begin{align}
	\bm{Y}_m = \mathcal{F}_m( \mathcal{S}_m(\bm{Z}_{m}) ), \forall m \in \bar{\mathcal{M}},
\end{align}
where $\mathcal{F}_m(\cdot)$ is the fully connected layer, $\mathcal{S}_m(\cdot)$ is the sliding-window pooling module, and $\bm{Y}_m \in \mathbb{R}^{s_{m} \times d_{\rm T}}$ is the resulting token with sequence length $s_{m}$ and dimension $d_{\rm T}$.
Subsequently, all token embeddings are reshaped and modulated to complex-valued signals for transmission over the wireless channel, which can be expressed as
\begin{align}
	\bm{y}_m=\mathcal{C}_m(\bm{Y}_m), \forall m \in \mathcal{M},
\end{align}
where $\mathcal{C}_m(\cdot)$ is the reshape and modulation module, and $\bm{y}_m \in \mathbb{C}^{\frac{s_m d_{\rm T}}{2}}$ is the resulting signal.
%$\bm{y}_m$ is constrained by a maximum power $P^{\rm max}_m$, i.e., $\frac{2}{s_m d_{\rm T}} \mathbb{E} \Vert \bm{y}_m \Vert^2 \le P^{\rm max}_m$.

\subsection{Token Receiver}
To avoid introducing interference among multimodal signals, we divide the system channel into $m$ frequency-orthogonal subchannels and adopt orthogonal frequency division modulation (OFDM) transmission. Thus, the received signal $\tilde{\boldsymbol{y}}_m$ can be expressed as
\begin{align}
	\tilde{\boldsymbol{y}}_m=h_m \sqrt{p_m} \boldsymbol{y}_m+\boldsymbol{n}_m, \forall m \in \mathcal{M},
\end{align}
where $h_m$ is the channel coefficient, $p_m$ is the transmit power and $\bm{n}_m$ is the channel noise vector which follows symmetric complex Gaussian distribution $\mathcal{CN}(0,\sigma^2\boldsymbol{I})$ with zero mean and variance $\sigma^2$.
Subsequently, the received signal $\tilde{\boldsymbol{y}}_m$ is demodulated and reshaped back into the token space,
\begin{align}
	\tilde{\bm{Y}}_m=\mathcal{C}_m^{-1}(\tilde{\bm{y}}_m), \forall m \in \mathcal{M},
\end{align}
where $\mathcal{C}_m^{-1}(\cdot)$ is the demodulation and reshape module. We adopt a projection head $\mathcal{G}_m(\cdot)$ consisting of two fully connected layers with a ReLU activation in between as the neural component of the token receiver \cite{simclr},
\begin{align}\label{received}
	\bm{Y}'_m=\mathcal{G}_m (\tilde{\bm{Y}}_m), \forall m \in \mathcal{M},
\end{align}
where $\bm{Y}'_{m} \in \mathbb{R}^{s_{m} \times d_{\rm T}}$. We concatenate $\bm{Y}'_{m}$ from all modalities $m \in \mathcal{M}$ as the input to the foundation model deployed on the server, which subsequently produces the task output $\bm{O}$,
\begin{align}
	\bm{O} = \mathcal{W}( \mathcal{U}(\boldsymbol{Y}_{1}^{\prime},\boldsymbol{Y}_{2}^{\prime},\ldots,\boldsymbol{Y}_{M-1}^{\prime}, \bm{Y}^{\prime}_{{\rm T}}) ),
\end{align}
where $\mathcal{W}(\cdot)$ is the foundation model and $\mathcal{U}(\cdot)$ is the concatenation module.
In task-oriented communication systems, the loss with respect to $\bm{O}$ is typically used directly to optimize the entire system \cite{wanting-2}. However, to exploit multimodal information more effectively, we divide the training into two stages, namely cross-modal alignment and task-oriented fine-tuning. The cross-modal alignment stage is designed to map multimodal token embeddings into a unified feature space, thereby enabling the foundation model to leverage multimodal information more efficiently. Details of the two-stage training algorithm are provided in Section~\ref{two-stage}, and its effectiveness is validated through simulations in Section~\ref{simulation}.

\section{Two-Stage Training Algorithm}\label{two-stage}
To strengthen the interaction of multimodal information and enhance the collaborative learning across modalities \cite{Align-then-refine}, we divide the training process into two stages: cross-modal alignment and task-oriented fine-tuning.

\subsection{Cross-Modal Alignment}
Since most foundation models are pretrained on text data, the cross-modal alignment aims to align token embeddings of other modalities with the text. To ensure the robustness of text token embeddings against noise, we introduce a reconstruction loss specifically for the received text token $\bm{Y}'_{\rm T}$ in \eqref{received}, which can be expressed as
\begin{align}\label{rec}
	\mathcal{L}^{\rm REC}_{\rm T} = \Vert \bm{Y}'_{\rm T}-\bm{Y}_{\rm T} \Vert^2.
\end{align}
We then introduce a contrastive loss on $\bm{Y}'_{m}$ to achieve cross-modal alignment. Since token sequences differ in length across modalities ($s_{m} \neq s_{m'}$ for $m, m' \in \mathcal{M}$ and $m \neq m'$), their pairwise similarity cannot be directly computed. Thus, we apply a pooling operation to extract a representative vector $\bm{a}_m \in \mathbb{R}^{d_{\rm T}}$ for each modality $m \in \mathcal{M}$.
Letting $\text{sim}(\boldsymbol{u},\boldsymbol{v})=\boldsymbol{u}^\top\boldsymbol{v}/\|\boldsymbol{u}\|\|\boldsymbol{v}\|$ denotes the cosine similarity, the contrastive loss for non-textual modalities can be expressed as
\begin{align}\label{con}
	\mathcal{L}^{\rm CON}_{m}\!=\!-\!\log\frac{\exp(\text{sim}(\bm{a}_{m},\bm{a}_{\rm T})/\tau)}{\sum_{m=1}^{M-1}\exp(\text{sim}(\bm{a}_{m},\bm{a}_{\rm T})/\tau)},\forall m \!\in\! \bar{\mathcal{M}},
\end{align}
where $\tau$ denotes the temperature parameter. 
Since the contrastive loss constrains the pooled vectors $\bm{a}_m$ rather than the full token sequences $\bm{Y}'_m$, it maps embeddings of different modalities into a unified feature space while preserving sequence-level details and modality-specific characteristics. During this stage, we train only the projection layers, with $\mathcal{F}_m(\cdot), \forall m \in \bar{\mathcal{M}}$ at the transmitter and $\mathcal{G}_m(\cdot), \forall m \in \mathcal{M}$ at the receiver, while keeping the modality encoders $\mathcal{E}_m(\cdot), \forall m \in \mathcal{M}$ and the foundation model $\mathcal{W}(\cdot)$ frozen.

\subsection{Task-Oriented Fine-Tuning}
After completing the cross-modal alignment stage, we fine-tune the entire system on the downstream task. Specifically, the multimodal token embeddings $\bm{Y}'_{m}$ in \eqref{received} are concatenated as $\hat{\bm{Y}} = \mathcal{U}(\boldsymbol{Y}_{1}^{\prime},\boldsymbol{Y}_{2}^{\prime},\ldots,\boldsymbol{Y}_{M-1}^{\prime}, \bm{Y}^{\prime}_{{\rm T}})$ and forwarded to the foundation model deployed on the server.
We consider the LoRA adapter for parameter-efficient fine-tuning \cite{lora}. For the foundation model $\bm{W}_0 \in \mathbb{R}^{d_{\rm T} \times k}$ with frozen parameters, we train its low-rank adapter $\Delta \bm{W} = \bm{B}\bm{A}^\top$, where $\bm{A} \in \mathbb{R}^{k \times r}$, $\bm{B} \in \mathbb{R}^{d_{\rm T} \times r}$ and $\mathrm{min} (d_{\rm T}, k) \gg r$. 
$\bm{A}_0$ and $\bm{B}_0$ are initialized with random values drawn from a zero-mean Gaussian distribution with a small standard deviation, ensuring that $\Delta \bm{W}$ remains close to zero at the start of training.
Specifically, the fine-tuned foundation model can be given by
\begin{align}
	\bm{W} = \bm{W}_0 + \Delta \bm{W} = \bm{W}_0 + \bm{B} \bm{A}^\top.
\end{align}
With the concatenated input token embedding $\hat{\bm{Y}}$, the frozen foundation model $\bm{W}_0$ and its trainable low-rank adapter $\Delta \bm{W}$, the entire system is fine-tuned using the following loss fuction to achieve task-oriented communication,
\begin{align}
	\mathcal{L}^{\rm Task} = \ell (\hat{\bm{Y}}; \bm{W}_0, \Delta \bm{W} ),
\end{align}
where $\ell(\cdot)$ is a task-related loss, e.g., cross-entropy loss.

\section{Resource Allocation}
Longer tokens can carry more information and improve model performance, but they also increase communication latency in resource-constrained networks.
Since text tokens are more compact and exhibit weaker robustness to errors, we allocate dedicated resources to them. 
For the tokens of other modalities, we formulate the optimization problem that balances transmission latency and model performance by jointly optimizing the bandwidth, power allocation vector and token length. The transmission latency and model performance are modeled as follows.

\subsection{Problem Formulation}
\subsubsection{Transmission Latency}
To avoid inter-modal signal interference, we allocate orthogonal subchannels $B_m$ to each device $m \in \bar{\mathcal{M}}$ within the channels excluding those used for transmitting textual tokens.
The uplink transmission rate of device $m$ is
\begin{align}
	R_m = B_m \log_2 (1+ \frac{h_m p_m}{N_0 B_m}), \forall m \in \bar{\mathcal{M}}.
\end{align}
where $N_0$ is the noise power. The transmission latency for $s_m$ token vectors can be expressed as
\begin{align}\label{latency}
	T_m  = \frac{s_m q}{B_m \log_2 (1+ \frac{h_m p_m}{N_0 B_m})}, \forall m \in \bar{\mathcal{M}},
\end{align}
where $q$ is the number of bits for each token vector.

\subsubsection{Model Performance}
%\indent{\color{blue} \textit{2) Inference Performance:}}
Due to the inexplicability of neural networks, the relationship between token length and model performance cannot be expressed in closed form. Therefore, we approximate this relationship using the statistics of model evaluations. Following \cite{meta}, the relation between token length and validation loss can be estimated by the following exponential model:
\begin{align}\label{performance}
	\phi_m(s_m) = (\frac{\alpha_m}{s_m})^{\beta_m} + \gamma_m, \forall m \in \bar{\mathcal{M}},
\end{align}
where $\alpha_m, \beta_m, \gamma_m \geq 0$ are tuning parameters, which can be obtained by curve fitting in practical implementions.

\subsubsection{Problem Formulation}
Both latency and model performance are important for the practical implementation of task-oriented token transmission in resource-constrained networks, thus it is desirable to minimize the transmission latency in \eqref{latency} and the validation loss in \eqref{performance}. However, it is difficult to minimize both metrics simultaneously. For example, increasing the token length helps reduce validation loss but inevitably increases transmission latency. Therefore, to strike a trade-off between communication and model performance, we formulate a weighted-sum optimization problem as follows
\begin{subequations}
	\vspace{-3mm}
	\label{eq:optim}
	\begin{align}
		\underset{\bm{B}, \bm{p}, \bm{s}}{\min} \;
		& (1-\lambda) \max_{m \in \bar{\mathcal{M}}} T_m
		+ \lambda \sum\nolimits_{m=1}^{M-1} \phi_m(s_m) \label{cost} \\
		\text{s.t.} \;
		& \sum\nolimits_{m=1}^{M-1} B_m \le B^{\rm max}, \label{bmax} \\
		& \sum\nolimits_{m=1}^{M-1} p_m \le P^{\rm max}, \label{pmax} \\
		& 0 \le B_m \le B^{\rm max}_m, \quad \forall m \in \bar{\mathcal{M}}, \label{b-range} \\
		& 0 \le p_m \le P^{\rm max}_m, \quad \forall m \in \bar{\mathcal{M}}, \label{p-range} \\
		& 0 \le s_m \le S_{m}^{\rm max}, \quad \forall m \in \bar{\mathcal{M}}. \label{smax}
	\end{align}
\end{subequations}
where $\bm{B}=[B_1, B_2,\ldots, B_{M-1}]^\top$ is the bandwidth allocation vector, $\bm{p}=[p_1, p_2,\ldots, p_{M-1}]^\top$ is the power allocation vector, $\bm{s}=[s_1, s_2,\ldots,s_{M-1}]^\top$ is the token length vector and $\lambda \in [0,1]$ is the weight to balance two metrics. Constraints \eqref{bmax} and \eqref{pmax} ensure that the allocated bandwidth and power do not exceed the total bandwidth and power budgets, respectively. Constraint \eqref{b-range}–\eqref{smax} guarantees that the bandwidth, power, and token length assigned to each device or modality remain within their respective feasible regions.

\vspace{-3mm}
\subsection{Proposed Solution}
The original problem involves strong coupling between $(\bm{B}, \bm{p}, \bm{s})$ across users and a max-delay term.
We first linearize the non-convex $\max$ operation via an auxiliary variable $T$:
\begin{subequations}
	\label{eq:optim1}
	\begin{align}
		\underset{\bm{s}, \bm{B}, \bm{p}, T}{\min}
		& \quad (1-\lambda) T+ \lambda \sum\nolimits_{m=1}^{M-1} \phi_m(s_m)    \label{cost1}\\
		\text{s.t.}
		& \quad \frac{s_m q}{B_m \log_2 (1+ \frac{h_m p_m}{N_0 B_m})} \le T, \forall m \in \bar{\mathcal{M}}, \label{aux-latency}\\
		& \quad \eqref{bmax}-\eqref{smax}.
	\end{align}
\end{subequations}
To solve the non-convex problem, we decouple it into to two subproblems and solve them iteratively.

\subsubsection{Optimization of token length}
Given the bandwidth $\bm{B}$ and power $\bm{p}$, problem \eqref{eq:optim1} can be rewritten as
\begin{subequations}
	\label{eq:optim2}
	\begin{align}
		\underset{\bm{s}, T}{\min}
		& \quad (1-\lambda) T+ \lambda \sum\nolimits_{m=1}^{M-1} \phi_m(s_m)    \label{cost2}\\
		\text{s.t.}
		& \quad \eqref{smax}~{\rm and}~\eqref{aux-latency}.
	\end{align}
\end{subequations}
It is a convex problem, and the closed-form optimal solution of $\bm{s}$ can be characterized.
Specifically, for each device $m$, the optimal token length satisfies
\begin{align}
	s_m^*=\begin{cases}
		\frac{T R_m}{q},  &\mathrm{if}~m\in \mathcal{A},\\
		S_m^{\max},  &\text{otherwise},
	\end{cases}
\end{align}
where $\mathcal{A} \subseteq \bar{\mathcal{M}}$ denotes the set of devices whose token lengths do not reach the maximum limit $S^{\rm max}_m$. The auxiliary variable $T$ is determined by solving the following scalar equation:
\begin{align}
	\sum\nolimits_{m \in \mathcal{A}}
	\lambda\beta_m\alpha_m^{\beta_m}\frac{q^{\beta_m}}{T^{\beta_m+1}R_m^{\beta_m}}=1-\lambda.
\end{align}
The above equation is continuous and strictly decreasing with respect to $T>0$, thus guaranteeing the existence and uniqueness of a feasible solution. The optimal $T$ can be obtained numerically, after which the optimal token length vector $\bm{s}^*$ follows directly.

\subsubsection{Optimization of power and bandwidth allocation}
Given the token length allocation $\bm{s}$, the optimization of bandwidth $\bm{B}$ and transmit power $\bm{p}$ can be formulated to find the minimum achievable latency. We can accomplish this by performing a bisection search over a candidate latency $T'$, which is to find the smallest $T^{\prime *}$ for which a feasible allocation of $\bm{B}$ and $\bm{p}$ exists. The search begins with an initial interval of $[0, T]$, where $T$ is the auxiliary variable obtained by solving the previous subproblem. At each bisection iteration, we check the feasibility of the current candidate latency $T'$, which involves solving the following feasibility problem:
\begin{subequations}
	\label{eq:optim3}
	\begin{align}
		{\rm find}
		& \quad \{\bm{B},\bm{p}\}  \\
		\text{s.t.}
		& \quad \frac{s_m q}{B_m \log_2 (1+ \frac{h_m p_m}{N_0 B_m})} \le T', \forall m \in \bar{\mathcal{M}}, \label{aux-T-2} \\
		& \quad \eqref{bmax}-\eqref{p-range}.
	\end{align}
\end{subequations}

The feasibility problem \eqref{eq:optim3} can be be formulated as a minimization problem to find the minimum required total transmit power, which is given by
\begin{subequations}
	\label{eq:optim4}
	\begin{align}
		\underset{\bm{B}, \bm{p}}{\min}
		& \quad  \sum\nolimits_{m=1}^{M-1} p_m \\
		\text{s.t.}
		& \quad \eqref{bmax}-\eqref{p-range}~{\rm and}~\eqref{aux-T-2}.
	\end{align}
\end{subequations}

Problem \eqref{eq:optim4} is non-convex. However, it can be converted into a convex optimization problem by first solving for the minimum power $p_m$ required for a given bandwidth $B_m$ to meet the latency constraint \eqref{aux-T-2}. Specifically, from \eqref{aux-T-2}, we have
\begin{align}
	p_m \geq \frac{N_0B_m}{h_m}\left(2^{\frac{s_m q}{B_mT^{\prime}}}-1\right), \forall m \in \bar{\mathcal{M}}.
\end{align}
To minimize power, we set $p_m$ to its minimum value
\begin{align}\label{mini-pm}
	p_m = \frac{N_0B_m}{h_m}\left(2^{\frac{s_m q}{B_mT^{\prime}}}-1\right), \forall m \in \bar{\mathcal{M}}.
\end{align}
By substituting \eqref{mini-pm} back into the objective function, the problem is reduced to a optimization problem over $B_m$, which is given by
\begin{subequations}
	\label{eq:optim5}
	\begin{align}
		\underset{\bm{B}}{\min}
		& \quad \sum\nolimits_{m=1}^{M-1} \frac{N_0B_m}{h_m}\left(2^{\frac{s_m q}{B_mT^{\prime}}}-1\right) \\
		\text{s.t.}
		& \quad \frac{N_0B_m}{h_m}\left(2^{\frac{s_m q}{B_mT^{\prime}}}-1\right) \le P^{\rm max}_m, \forall m \in \bar{\mathcal{M}}, \\
		& \quad \eqref{bmax}~{\rm and}~\eqref{b-range}.
	\end{align}
\end{subequations}

Problem \eqref{eq:optim5} is convex and can be solved numerically. Thus, The feasibility prolem \eqref{eq:optim3} of $T'$ is then determined by comparing the resulting minimum total power with the available power budget $P^{\rm max}$.
Let $\mathcal{C}_{T'}$ denote the set of feasible $\{\bm{B}, \bm{p}\}$. If $\mathcal{C}_{T'}$ is non-empty, the current $T'$ is feasible and the upper bound is updated to $T'$. Otherwise, the current $T'$ is infeasible and the lower bound is updated to $T'$.
The bisection continues until the interval width falls below a predefined tolerance $\epsilon$, yielding the minimum achievable latency $T^{\prime*}$ and the corresponding feasible allocation $\{\bm{B}^{*}, \bm{p}^{*}\}$.

\section{Simulation Results}\label{simulation}
\begin{table}[t]
	\centering
	\renewcommand{\arraystretch}{1.4}
	\caption{Simulation Parameters}
	\label{sim-para}
	\begin{tabular}{|l|l|l|l|}
		\hline
		\textbf{Parameter}         & \textbf{Value} & \textbf{Parameter}        & \textbf{Value}            \\ \hline
		$B^{\rm max}$              & $3$ MHz        & $P^{\rm max}$             & $23$ dBm                  \\ \hline
		$N_0$                      & $-174$ dBM/Hz  & ${\rm PL}(d)$             & $128.1\!+\!37.6 \log_{10}(d)$ \\ \hline
		$d$                        & $[0.3, 0.5]$ km & $\lambda$                & $0.6$                    \\ \hline
		$S_{\rm visual}^{\rm max}$ & $128$          & $S_{\rm audio}^{\rm max}$ & $64$                     \\ \hline
	\end{tabular}
	\vspace{-3mm}
\end{table}
In the simulations, we consider a multimodal communication network involving textual, audio, and visual modalities. Text tokens are allocated dedicated resources with bandwidth $B_{\rm T}=0.5$ MHz and power $p_{\rm T} = 15$ dBm. Other parameters are shown in Table~\ref{sim-para}.
The foundation model used is Qwen2.5-1.5B \cite{qwen} with a low-rank adapter $r=4$, deployed on the edge server. Modality-specific encoders are: Qwen text encoder for text, AST \cite{ast} for audio, and ViViT \cite{vivit} for video, all deployed on edge devices.
We perform cross-modal alignment on the VALOR-32K dataset \cite{valor} and fine-tune the model for audio-visual question answering (AVQA) on the MUSIC-AVQA dataset \cite{music}. 
Let $N$ denote the total number of test samples and $\mathbf{1}$ be the indicator function. For a predicted word $\acute{o}_i$ and the corresponding ground-truth word $o_i$, the accuracy is computed as $\frac{1}{N}\sum_{i=1}^N \mathbf{1}\{\acute{o}_i=o_i\}$. The following schemes are considered as benchmarks:
1) Proportional bandwidth and water-filling power allocation (PB-WF).
2) Equal resource allocation (ERA).
3) Ideal token communication (ITC).
4) Token communication without alignment (TC-NA).
5) Token communication with half of maximum length tokens (TC-HL).

\begin{figure}[t]
	\centering
	% 第一行
	\begin{minipage}{0.24\textwidth}
		\centering
		\includegraphics[scale=0.47]{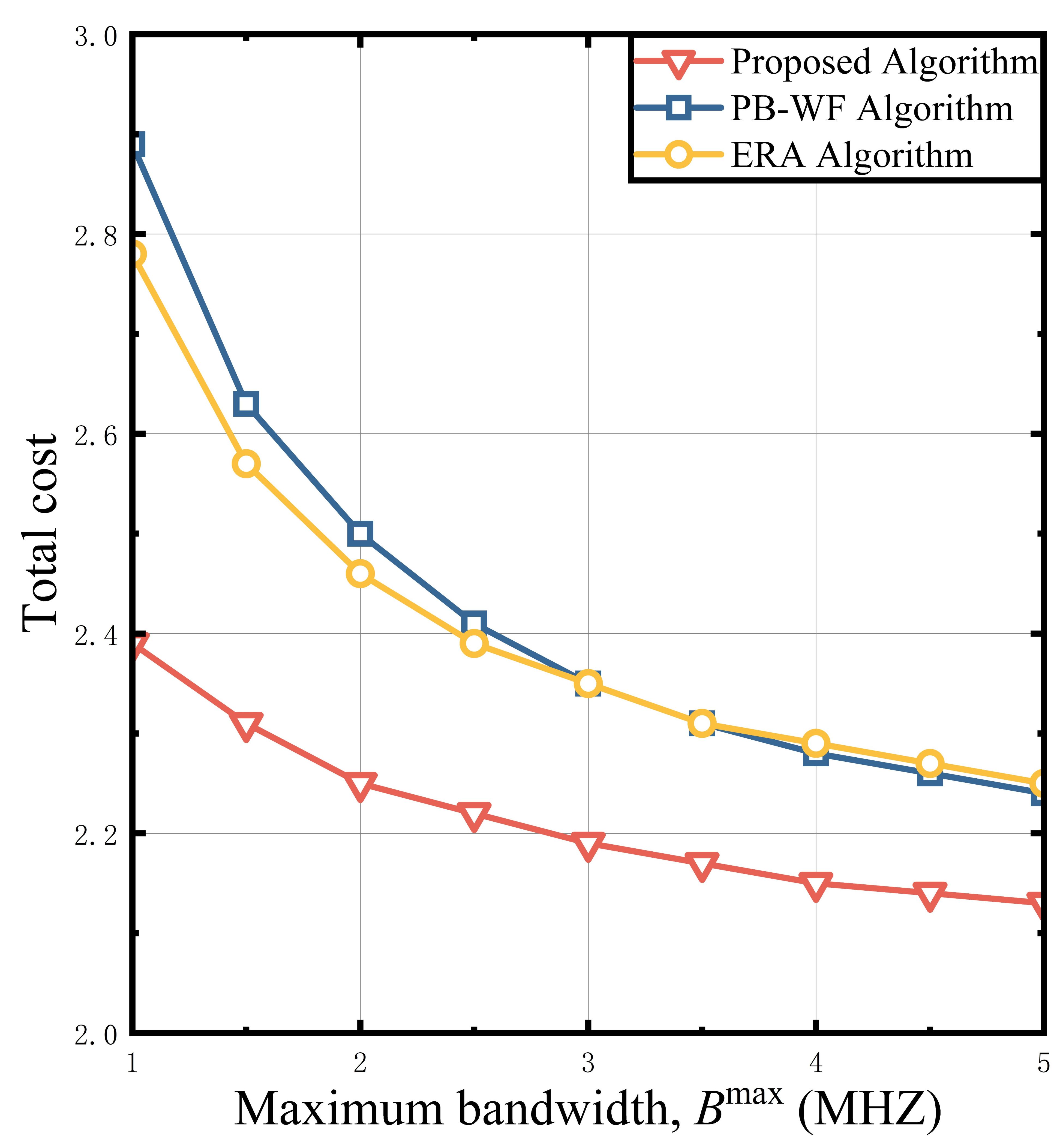}
		\caption{Total cost vs. $B^{\rm max}$}
		\label{b-cost}
	\end{minipage}
	\hfill
	\begin{minipage}{0.24\textwidth}
		\centering
		\includegraphics[scale=0.47]{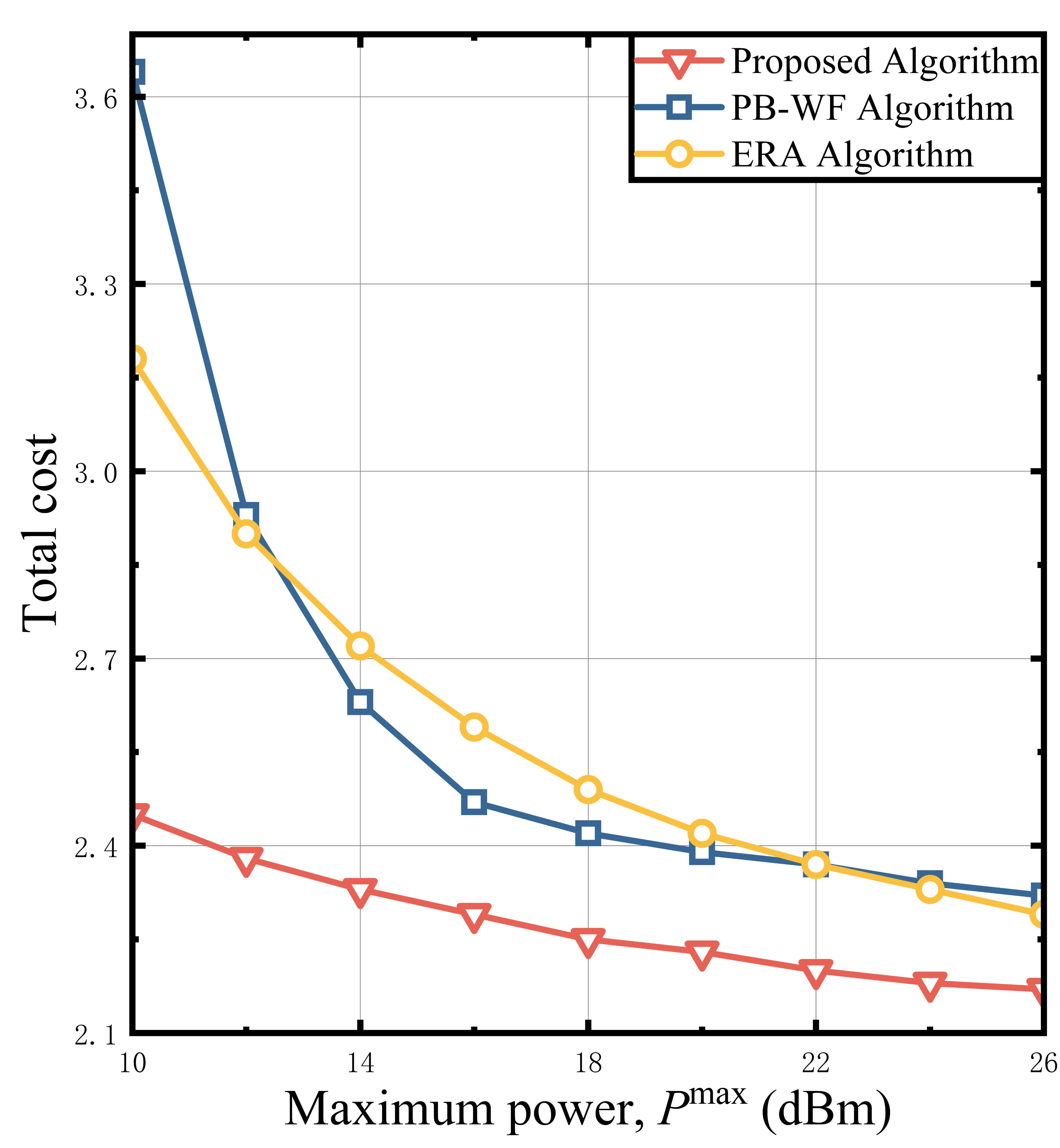}
		\caption{Total cost vs. $P^{\rm max}$}
		\label{p-cost}
	\end{minipage}
	
%	\vspace{0.7em}
	
	\begin{minipage}{0.24\textwidth}
		\centering
		\includegraphics[scale=0.47]{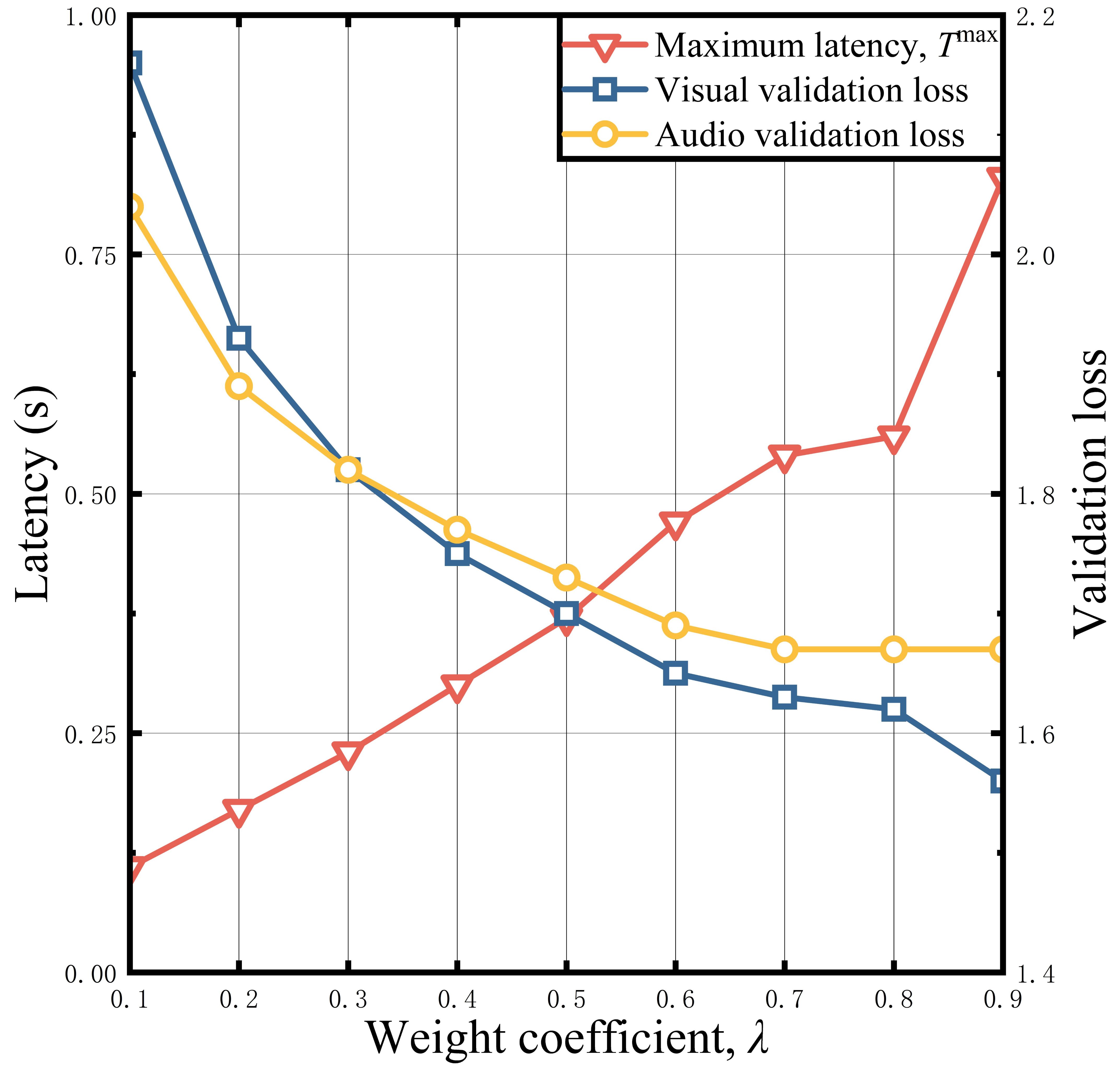}
		\caption{Impact of $\lambda$}
		\label{lambda}
	\end{minipage}
	\hfill
	\begin{minipage}{0.24\textwidth}
		\centering
		\includegraphics[scale=0.47]{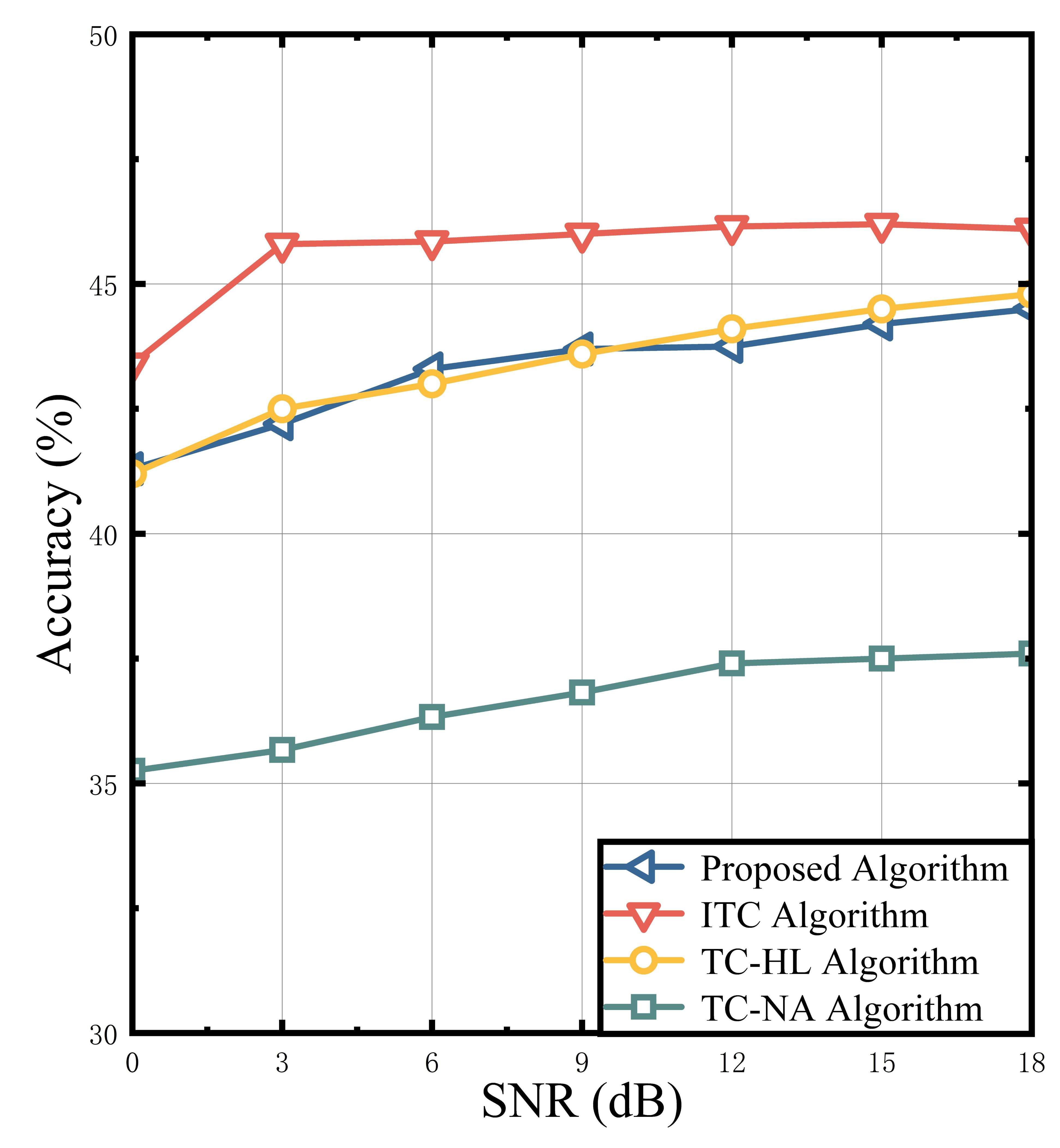}
		\caption{Accuracy vs. SNR}
		\label{SNR}
	\end{minipage}
	\vspace{-3mm}
\end{figure}

Figs.~\ref{b-cost} and \ref{p-cost} present the impact of maximum bandwidth and power budget on the total cost defined in \eqref{eq:optim}, respectively. As the available bandwidth and power increase, the system is able to transmit more tokens, which enhances model performance without causing a significant increase in latency. Furthermore, the proposed solution consistently outperforms both the PB-WF and ERA algorithms across different settings. Fig.~\ref{lambda} shows that when $\lambda$ is larger, the system emphasizes model performance, which leads to higher latency but lower validation loss.
Fig.~\ref{SNR} presents test accuracy under varying SNRs. The proposed algorithm achieves results comparable to TC-HL, while test accuracy remains significantly higher than TC-NA even with 
shorter tokens. Simulation results demonstrate that the two-stage training algorithm achieves higher performance than the version without cross-modal alignment, even with fewer tokens. Moreover, for SNRs above $3$ dB, the performance of maximum token length fluctuates less than $0.5\%$, while half the length achieves a $2.3\%$ gain at high SNRs, which implies that token usage improves robustness.

\section{Conclusion}
	In this letter, we propose a task-oriented multimodal token transmission scheme and employ a two-stage training algorithm to enhance inter-modal consistency and task-relevant token transmission. By jointly optimizing bandwidth, power, and token length, the framework balances resource usage and model performance. Simulation results demonstrate that our proposed scheme can reduce total cost while maintaining model performance.
%\newpage
%\appendices

\bibliographystyle{IEEEtran}
\bibliography{IEEEabrv}

\end{document}